\documentclass[twocolumn,aps,prd,superscriptaddress,nofootinbib,floatfix]{revtex4-2}

\usepackage{graphicx}
\usepackage{multirow}
\usepackage[colorlinks=true,citecolor=blue,urlcolor=blue,linkcolor=blue]{hyperref}

\usepackage{amssymb}
\usepackage{amsmath}
\usepackage{xcolor,soul}
\usepackage{siunitx,xspace,wasysym}
\usepackage[normalem]{ulem}

\newcommand{\nn}{\nonumber}
\newcommand{\beq}{\begin{equation}}
\newcommand{\eeq}{\end{equation}}
\newcommand{\beqa}{\begin{eqnarray}}
\newcommand{\eeqa}{\end{eqnarray}}

\newcommand{\GeV}{{\rm GeV}}
\newcommand{\MeV}{{\rm MeV}}

\def\lqcd{\Lambda_{\rm QCD}}
\newcommand{\ds}{\displaystyle}

\newcommand{\ups}[1]{$\Upsilon(#1S)$}
\newcommand{\Bbar}{\,\overline{\!B}{}}

\newcommand{\Dbar}{\,\overline{\!D}{}}
\newcommand{\Kbar}{\,\overline{\!K}{}}
\def\B0bar{\Bbar{}^0}
\def\D0bar{\Dbar{}^0}
\def\K0bar{\Kbar{}^0}

\newcommand{\Rpmz}{\ensuremath{R^{\pm0}}\xspace}
\newcommand{\fnB}{\ensuremath{f_{\not B}}\xspace}

\def\OMIT#1{{}}
\makeatletter
\g@addto@macro\bfseries{\boldmath}
\let\Hy@backout\@gobble
\makeatother

\begin{document}

\title{Novel approaches to determine \texorpdfstring{$B^\pm$}{B+-} and \texorpdfstring{$B^0$}{B0} meson production fractions}

\author{Florian Bernlochner}
\affiliation{Physikalisches Institut der Rheinischen Friedrich-Wilhelms-Universit\"at Bonn, 53115 Bonn, Germany}

\author{Martin Jung}
\affiliation{Dipartimento di Fisica, Universit\`a di Torino \& INFN, Sezione di Torino, I-10125 Torino, Italy}

\author{Munira Khan}
\affiliation{Physikalisches Institut der Rheinischen Friedrich-Wilhelms-Universit\"at Bonn, 53115 Bonn, Germany}

\author{Greg Landsberg}
\affiliation{Brown University, Dept.\ of Physics, Providence, RI 02912, USA}

\author{Zoltan Ligeti}
\affiliation{\mbox{Ernest Orlando Lawrence Berkeley National Laboratory,
University of California, Berkeley, CA 94720, USA}}

\begin{abstract}

We propose novel methods to determine the $\Upsilon(4S)\to B^+B^-$ and $\Upsilon(4S)\to B^0\B0bar$ decay rates. 
The precision to which they and their ratio are known yields at present a limiting uncertainty around $2\%$ in measurements of absolute $B$ decay rates, and thus in a variety of applications, such as precision determinations of elements of the Cabibbo--Kobayashi--Maskawa matrix and flavor symmetry relations. The new methods we propose are based in one case on exploiting the $\Upsilon(5S)$ data sets, in the other case on the different average number of charged tracks in $B^\pm$ and $B^0$ decays.  We estimate future sensitivities using these methods and discuss possible measurements of $f_d / f_u$ at the (HL-)LHC.

\end{abstract}

\maketitle

\section{Introduction}

Precise knowledge of the absolute branching fractions of charged and neutral $B$ meson decays is crucial for a large part of the flavor physics program, spanning the range from understanding hadronic physics to new-physics searches. 
They enter precision determinations of fundamental Standard Model (SM) parameters, such as 
elements of the Cabibbo--Kobayashi--Maskawa (CKM) matrix, and flavor symmetry relations, which in turn impact the sensitivity of $CP$-violation measurements to physics beyond the SM.
As Belle and BaBar recorded hundreds of millions of $B$ meson decays, projected to increase by nearly two orders of magnitude at Belle~II, a significant uncertainty in the otherwise ever more precise measurements has become the ratio of $\Upsilon(4S)$ decay to charged vs.\ neutral $B$ mesons,
\beq\label{Rpmzdef}
\Rpmz = \frac{\Gamma(\Upsilon(4S)\to B^+B^-)} {\Gamma(\Upsilon(4S)\to B^0\B0bar)} \,. 
\eeq

Although the mass difference $m_{B^0} - m_{B^+} = (0.32\pm 0.05)\,\MeV$~\cite{Workman:2022ynf} is small, the restricted phase space in the $\Upsilon(4S) \to B\Bbar$ decay of merely 20\,MeV and the resulting small velocity of the $B$ mesons give rise to enhanced electromagnetic effects and isospin violation~\cite{Atwood:1989em}.  
The range of theoretical predictions for these effects is substantial, spanning \Rpmz values beyond 1\,--\,1.2~\cite{Atwood:1989em, Lepage:1990hm, Byers:1990rd, Kaiser:2002bm, Voloshin:2003gm, Voloshin:2004nu, Dubynskiy:2007xw, Milstein:2021fnc}, much above the desired precision.
Therefore, direct experimental determinations are crucial.  The world average obtained by HFLAV~\cite{Amhis:2022mac} is
\beq\label{eq::PDG}
\Rpmz = 1.057^{+0.024}_{-0.025}\,,
\eeq
indicating a notable deviation from unity, albeit smaller than predicted by some theoretical estimates.

Given the above range of applications, reducing this uncertainty to the (sub-)percent level would be very important
(as discussed, e.g., in Refs.~\cite{Jung:2012mp,Grossman:2012ry,Jung:2014jfa,Ligeti:2015yma, Jung:2015yma}.)
An intrinsic challenge of such a determination is the difficulty of separating the 
production fractions from $B$ meson decay rates, since the most often measured quantities determine only their product.
Moreover, the measurements usually assume that $\Upsilon(4S)$ decays exclusively to $B$ meson pairs.  What is meant by this, is that the enhancement of the total $e^+e^-$ cross section near the $\Upsilon(4S)$ resonance equals (within a few permille) the $B$ meson production rate, with $B$ meson production kinematically forbidden for $\sqrt s \lesssim m_{\Upsilon(4S)} - \Gamma_{\Upsilon(4S)}$.  We denote $f_\pm = \Gamma(\Upsilon(4S)\to B^+B^-) / \Gamma_{\Upsilon(4S)}$, $f_{00} = \Gamma(\Upsilon(4S)\to B^0\B0bar) / \Gamma_{\Upsilon(4S)}$, and $\fnB = 1 - f_\pm - f_{00}$.  
Clearly, $\Rpmz = f_\pm/f_{00}$, so to relate \Rpmz to absolute branching fractions, knowledge of \fnB is required.  A lower bound on \fnB is obtained from the sum of measured $\Upsilon(4S)$ decays to lighter bottomonia and pions~\cite{Amhis:2022mac},
\beq\label{BRnonB}
\fnB > (0.264 \pm 0.021)\, \% \,.
\eeq
The strongest constraint, not yet included by HFLAV, is from CLEO~\cite{CLEO:1995umr},
\beq\label{BRnonB2}
\fnB = (-0.11 \pm 1.43 \pm 1.07)\,\% \,,
\eeq
with a larger uncertainty than the desired precision.

Another implicit assumption when averaging the available measurements is that the center-of-mass energy and the beam energy spread at which the $\Upsilon(4S)$ are produced, are similar at the relevant colliders and data-taking periods. This will be further discussed in Sec.~\ref{sec:discussion}.

In this paper we propose new methods to address these challenges, both in the short term and in the long term.
In Sec.~\ref{sec:reapp}, we reappraise the theoretical assumptions in various \Rpmz measurements, and update its world average.  
In Sec.~\ref{sec:5Sidea}, we propose a new method to determine \Rpmz with small theoretical uncertainties, using $\Upsilon(5S)$ data anticipated at Belle~II in the next decade.  
In Sec.~\ref{sec:tracks}, we propose a new method based on the different average number of charged-particle tracks in charged and neutral $B$ decays.  
In Sec.~\ref{sec:LHC} we discuss possible measurements of the $B^0$ to $B^+$ meson production ratio $f_d / f_u$ at the \mbox{(HL-)LHC}. 
Finally, Sec.~\ref{sec:discussion} discusses additional issues related to collider running conditions and concludes.

\begin{table*}
\renewcommand{\arraystretch}{1.2}
\begin{tabular}{llll}\hline\hline
\multicolumn{1}{c}{$\Rpmz$} & Method  & Comment & Reference\\
\hline\hline
1.047(44)(36) & Single vs.\ double-tag & Uses $f_{\not B}$, see text & \cite{BaBar:2005uwr,CLEO:1995umr,Amhis:2022mac}\\
\hline
1.039(31)(50) & $B\to X_c\ell\nu$ & Assumes negligible isospin violation & \cite{Hastings:2002yk,Belle:2002lms}\\
1.068(32)(20)(21) & $B\to X_s\gamma$ & Third uncertainty due to resolved photon contributions & \cite{Belle:2018iff}\\
\hline
1.055(30)  & & Average categories I and II &\\
\hline
1.065(12)(19)(32) & $B\to J/\psi K$ & Third uncertainty due to isospin violation in $B\to J/\psi K$ & \cite{Belle:2022hka,BELLE:2019xld}\\
1.013(36)(27)(30) & $B\to J/\psi K$ & Third uncertainty due to isospin violation in $B\to J/\psi K$ & \cite{BaBar:2004igf}\\
1.100(35)(35)(33) & $B\to J/\psi(ee) K$ & Third uncertainty due to isospin violation in $B\to J/\psi K$ & \cite{Belle-II:2022dbo}\\
1.066(32)(34)(32) & $B\to J/\psi(\mu\mu) K$ & Systematic uncertainties $\sim 100\%$ correlated with $ee$ mode & \cite{Belle-II:2022dbo}\\
\hline
1.060(18)(32)     & & Average for $B\to J/\psi K$ &\\
\hline
\textbf{1.057(23)}    & & {\bf Average of all categories I--III} &\\
\hline\hline
\end{tabular}
\caption{\label{tab::Rsummary} Available measurements for $\Rpmz$ from the three categories, as explained in detail in the main text.}
\end{table*}

\section{Present status of \texorpdfstring{\Rpmz}{R+-0}}
\label{sec:reapp}

In this section we update the analysis of Ref.~\cite{Jung:2015yma}, with the main difference that we allow for $\fnB\neq 0$.
In order to separately determine the production fractions and decay rates,
three categories of measurements have been commonly used so far (which we label I, II, and III below): 
\paragraph*{\bf I~~Cancellation of final-state dependence.} 
    This technique relies on the observation that for double-tagged events in $\Upsilon(4S)$ decays, the $B^+B^-$ and $B^0\B0bar$ production fractions enter linearly, while the decay rate enters quadratically (a technique developed for $\psi(3770)$~\cite{MARK-III:1985hbd}). This allows for a cancellation of the dependence on the decay rates in the ratio of the number of single-tag events squared and the number of double-tag events, while retaining that on the production fractions, thus making a theoretically clean measurement of isospin violation in production possible. 
\paragraph*{\bf II~~Known ratio of decay rates.}
    Considering any ratio of a charged to a neutral $B$ meson decay, the experimentally determined quantity is proportional to $\Rpmz$ times the ratio of the corresponding decay rates. If the ratio of decay rates is known, it is possible to extract the ratio \Rpmz. 
    Taking the ratio of decay rates from an external measurement relies on the determination of \Rpmz at the corresponding experiment, while for an extraction without external inputs, the knowledge of the ratio of decay rates has to stem from theory. Given the required level of precision, presently the only method available relies on isospin symmetry.
    While generally, a precise theoretical determination of isospin violation is extremely difficult, there are a few cases in which two decays are not only related by isospin symmetry, but isospin breaking is additionally suppressed. This is the case, e.g., for inclusive semileptonic $B$ meson decays, where the operator product expansion and heavy-quark symmetry provide an additional $\lqcd^2/m_{c,b}^2$ suppression~\cite{Chay:1990da} of the isospin breaking from both the strong interaction (as discussed, e.g., in Ref.~\cite{Gronau:2006ei}) and from electromagnetic effects. 
    In this case, isospin breaking can be safely assumed to be below~$1\%$.
\paragraph*{\bf III~~(Pseudo-)Isospin symmetry.} 
    Given the potential enhancement of isospin breaking in production, it is possible to extract it assuming that the breaking for (pseudo-)isospin-related decays is small compared to the one in production. We call \emph{pseudo-isospin relations} those in which the amplitudes are \emph{not} equal by isospin symmetry alone, but the unrelated contributions are expected to be of similar size as generic isospin breaking.
    This is the case, e.g., for $B\to J/\psi K$ decays, in which the annihilation amplitude contributing only to the charged mode is often argued to be negligible. The remaining isospin breaking is expected to be at the percent level. Clearly, making this assumption precludes the extraction of isospin violation in decay at the same order (and especially in the same decays for which this assumption has been made). This holds also for the values quoted in Eq.~\eqref{eq::PDG}, since some of the measurements in the average use this assumption. Furthermore, this strategy relies on the assumption that the isospin breaking in production is much larger than that in decay. While reasonable, this assumption is not firmly established experimentally yet, given that the result in Eq.~\eqref{eq::PDG} is only about two standard deviations from unity.

The available measurements of $\Rpmz$ are collected in Table~\ref{tab::Rsummary}.
The only measurement from category~I is the BaBar result (using about 82\,fb$^{-1}$ of data) for the production fraction of $B^0$ mesons~\cite{BaBar:2005uwr},
\beq
f_{00} = 0.487\pm0.010\pm0.008\,,
\eeq
where the dominant systematic uncertainty stems from the number of $B\Bbar$ pairs. To turn this into the determination of $\Rpmz$ in Table~\ref{tab::Rsummary}, information regarding the non-$B\Bbar$ fraction in $\Upsilon(4S)$ is required.

From category~II, Belle~\cite{Belle:2002lms} used inclusive semileptonic decays to measure $\Rpmz$. This result needs to be updated to the common ratio of $B$ meson lifetimes, the dominant systematic uncertainty in this case.%
\footnote{We follow the HFLAV procedure to account for the change in the central value and uncertainty of the lifetime ratio. Notably $\Rpmz$ and $\tau_{B^+}/\tau_{B^0}$ are anticorrelated~\cite{Hastings:2002yk}.}
It would also be interesting to revisit this analysis technique, where \Rpmz was determined simultaneously with $\Delta m_d$ in a self-consistent way. The data set used, \SI{30}{\per\femto\barn}, was only a small fraction of the full Belle or current Belle~II data.

Another measurement belonging to category~II is from the isospin asymmetry $A_I$ between $CP$-averaged rates in $B\to X_s\gamma$ decays,
\beq 
A_I(B\to X_s\gamma) \equiv \frac{\Gamma(\bar B^0\to X_s\gamma)-\Gamma(B^-\to X_s\gamma)}{\Gamma(\bar B^0\to X_s\gamma)+\Gamma(B^-\to X_s\gamma)}\,.
\eeq
While the so-called resolved photon contributions affect the isospin asymmetry, this effect is probably subdominant.  Therefore, this mode might not be suitable to achieve percent-level precision, but it is still useful, given the current measurements and uncertainties.  
Assuming isospin symmetry, except for a 2\% uncertainty due to the isospin-violating part of the resolved photon contributions~\cite{Bernlochner:2020jlt, Gunawardana:2019gep, Benzke:2010js}, the Belle measurement 
$A_I(B\to X_s\gamma)=(-0.48\pm1.49\pm0.97\pm1.15)\%$~\cite{Belle:2018iff} (using all Belle data) translates to the value listed in Table~\ref{tab::Rsummary}. 

Measurements in category~III are presently dominated by  $B\to J/\psi K$ decays. The experimentally determined quantity in these measurements (for a pair of pseudo-isospin-related final states, $F$) is the ratio
\beq
q_{F} \equiv \Rpmz\, \frac{{\cal B}(B^-\to F^-)}{{\cal B}(\bar B^0\to F^0)} 
= \Rpmz\, \frac{\tau_{B^-}}{\tau_{B^0}}\frac{\Gamma(B^-\to F^-)}{\Gamma(\bar B^0\to F^0)}\,.
\eeq
The $B^0$ and $B^\pm$ lifetimes, $\tau_{B^0}$ and $\tau_{B^-}$, respectively, are typically determined separately (but need to be used consistently when combining measurements), and \emph{either} the ratio of rates \emph{or} the ratio of production fractions can be determined, making an assumption about the other. The values in Table~\ref{tab::Rsummary} correspond to the assumption that the ratio of rates is equal to unity, and assigning a 3\% uncertainty to that assumption (as discussed below). Turning this around, using our averaged value for $\Rpmz$ based on the measurements from the first two categories, we obtain
\beq\label{eq::rGamma}
\frac{\Gamma(B^-\to J/\psi K^-)}{\Gamma(\B0bar\to J/\psi \bar K^0)} = 1.005 \pm 0.033 \,,
\eeq
or, equivalently,
\beq
A_I(B\to J/\psi K) = -0.002 \pm 0.017\,,
\eeq
where now the uncertainty due to the production fractions is taken into account consistently. This shows no indication of a sizable violation of the pseudo-isospin relation.

\begin{figure}[t!]
    \includegraphics[width=.9\columnwidth]{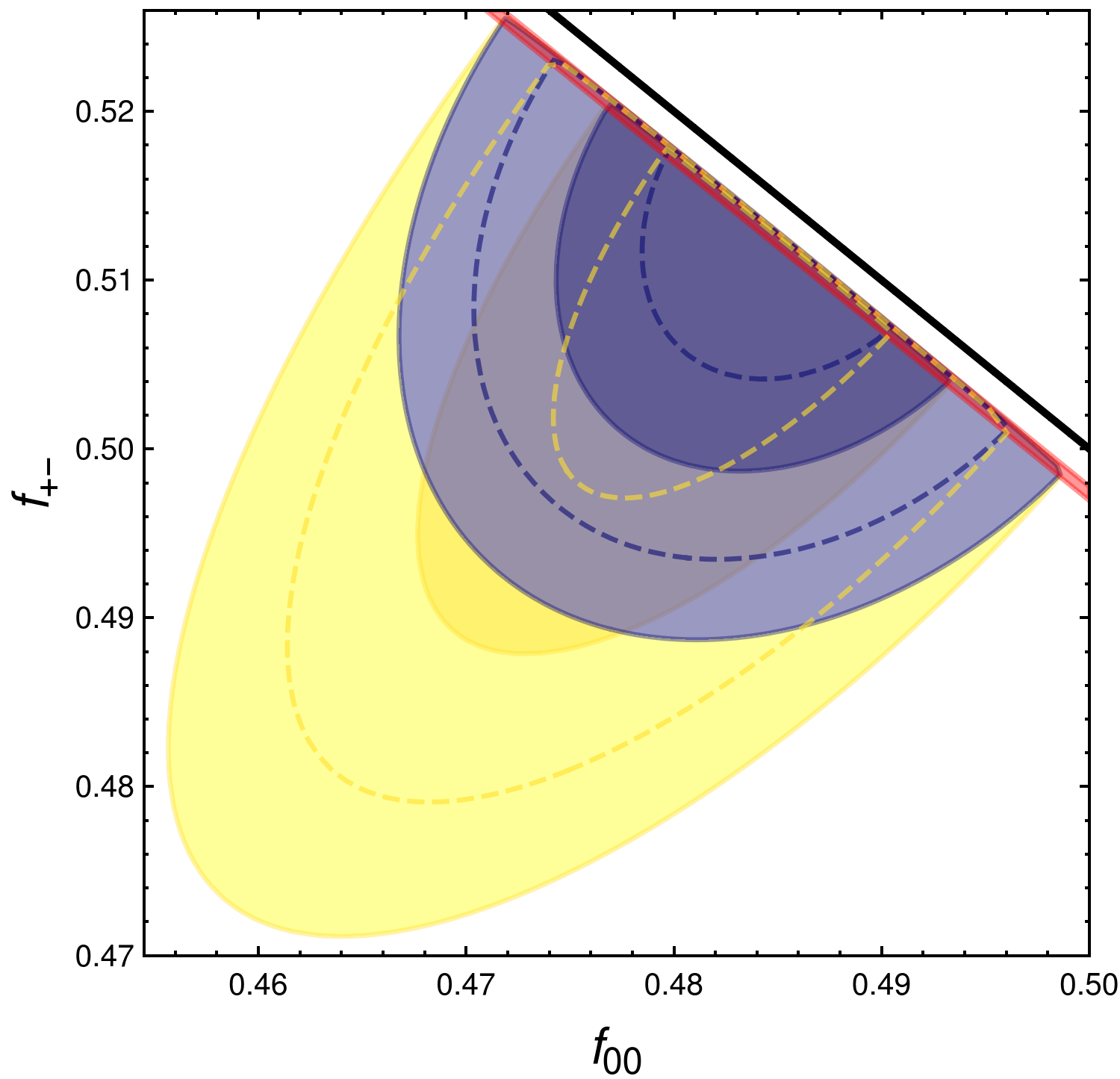}
    \caption{Impact of the treatment of \fnB on the determination of the $B\bar B$ production fractions. The black line corresponds to setting $\fnB = 0$, i.e., $f_{00}+f_\pm=1$. The red line corresponds to setting \fnB to the lower bound in Eq.~\eqref{BRnonB}. The yellow shaded areas use our results in Table~\ref{tab::Rsummary} and treat Eq.~\eqref{BRnonB} as a lower limit, while the blue shaded constraints include the CLEO measurement in Eq.~\eqref{BRnonB2}. The lighter and darker regions show $\Delta\chi^2\leq 5.99$ and $2.28$, respectively, while the dashed lines correspond to $\Delta \chi^2=1,4$ (illustrating one-dimensional limits).}
    \label{fig:f00fpm}
\end{figure}

A few comments regarding the values in Table~\ref{tab::Rsummary} are in order:
\begin{itemize}
    \item The measurements show excellent consistency, even among the different categories. While the uncertainties are sizable, this indicates that the isospin asymmetry is not anomalously large in the modes used for this determination, namely in $B\to J/\psi K$ decays.
    This is quantified in Eq.~\eqref{eq::rGamma}, which also motivates the uncertainty assigned to it above: isospin conservation was not assumed in obtaining Eq.~\eqref{eq::rGamma}, but is experimentally seen to hold at this level.
    \item The average of all the values in Table~\ref{tab::Rsummary}, including our estimates for the uncertainty due to isospin violation, results in 
    \beq\label{eq:Rpmzfull}
    \Rpmz_{\rm I+II+III} = 1.057 \pm 0.023\,,
    \eeq
    which is numerically close to the HFLAV average, but more robust, since it includes additional uncertainties for the assumptions made. The reason is that additional measurements are included here \cite{Belle-II:2022dbo, Belle:2022hka,CLEO:1995umr,Belle:2018iff}.
    \item
    In principle, \Rpmz and $f_{00}$ together determine \fnB via $\fnB=1-f_{00}(1+\Rpmz)=-0.003\pm0.029$; however, this uncertainty is still larger than that in Eq.~\eqref{BRnonB2}.
    \item
    Since we include \fnB in our calculations, the average for \Rpmz does not represent the full information from our analysis. Specifically, while with $\fnB=0$ the value for \Rpmz is in a one-to-one correspondence with $f_{00}$ and $f_\pm$, this no longer holds for $\fnB\neq 0$. For instance, determining $f_{\pm}$ now requires two of the measured quantities:
    \beqa
    f_\pm &=& f_{00}\Rpmz = 1-f_{00}-\fnB \nn\\
    &=& \frac{\Rpmz(1-\fnB)}{1+\Rpmz}\,.
    \eeqa
    However, it is $f_{00}$ and $f_\pm$ that determine the precision of absolute branching fractions, not \Rpmz. While our result for \Rpmz is numerically close to the one from HFLAV~\cite{Amhis:2022mac}, the results for the production fractions are quite different, since we include the CLEO measurement of \fnB \cite{CLEO:1995umr}. This is illustrated in Fig.~\ref{fig:f00fpm}, where we compare the impact of different treatments of \fnB on $f_{00}$ and $f_\pm$. This highlights the importance of determining \fnB with better precision. The resulting 
    uncertainties are asymmetric and highly correlated, and
    the central values and $\Delta\chi^2=1$ ($\Delta\chi^2=4$) ranges of the production fractions from the fit including the CLEO measurement are:
    \beqa
    f_\pm&=& 0.512\,\,[0.504,0.518]\,\ ([0.493,0.523])\,, \nonumber\\
    f_{00}&=&0.485\,\,[0.478,0.491]\,\ ([0.470,0.496])\,,\nonumber\\ 
    \fnB&=&0.003\,\,[0.002,0.014]\,\ ([0.002,0.029])\,.\label{eq:fpm00full}
    \eeqa
    The fit results without the CLEO measurement are,
    \beqa
    f_\pm&=& 0.512\,\,[0.497,0.518]\,\ ([0.479,0.523])\,, \nonumber\\
    f_{00}&=&0.485\,\,[0.474,0.491]\,\ ([0.461,0.496])\,,\nonumber\\ 
    \fnB&=&0.003\,\,[0.002,0.027]\,\ ([0.002,0.056])\,,
    \eeqa
    so our analysis reduces the uncertainties in $f_{\pm}$ and $f_{00}$ from about $2\%$ to $1.5\%$. The difference
    would be even larger without the measurement of $f_{00}$; this shows again the necessity to determine individual production fractions for either $B$ mesons or non-$B\bar B$ states.
\end{itemize}

Interestingly, the values in Table~\ref{tab::Rsummary} are not only consistent with one another, but also with the value obtained considering only the phase-space difference between $\Upsilon(4S)\to B^0\B0bar$ and $\Upsilon(4S)\to B^+ B^-$,
\beq \label{eq:rp_ps}
\Rpmz_{\mathrm{PS}} = \frac{p_{\pm}^3}{p_0^3} \approx 1.048\,.
\eeq
This value is larger than may be naively expected, due to the small phase space, which amplifies the impact of the small mass difference between the charged and neutral $B$ mesons.  On the other hand, the naive Coulomb enhancement of the charged mode, in the nonrelativistic limit and assuming point-like mesons is~\cite{Atwood:1989em}:
\beq \label{eq:rp_qed}
\Rpmz_{\mathrm{CE}} = \frac{2\pi\lambda(1+\lambda^2)}{1-\exp(-2\pi\lambda)}\,,
\eeq
where $\lambda = \alpha/(2v_\pm)$ denotes the Coulomb parameter (and $v_\pm = (1-4m_{B^{(*)\pm}}^2/m_\Upsilon^2)^{1/2}$, as appropriate for the $B$ or $B^*$ states in the $\Upsilon(4S)$ or $\Upsilon(5S)$ decays), which yields the values in the second last column of Table~\ref{tab::R0values}.%
\footnote{Since $m_{B^0} > m_{B^+}$, in $\Upsilon(nS) \to B \Bbar$, the phase space difference and the Coulomb enhancement of the charged mode go in the same direction.  (This also holds for $\Upsilon(5S) \to B^* \Bbar^*$ discussed below, though in that case the phase space effect is very small.)} Evidently, the large enhancement expected from this estimate is reduced, given that
\beq
\Rpmz_{\rm I+II+III}/\Rpmz_{\mathrm{PS}} = 1.008\pm0.022\,,
\eeq
consistent with small isospin violation beyond the phase space factor.
Nevertheless, the (additional) production asymmetry from isospin violation in the $\Upsilon(4S)$ decay may still be larger than without any enhancement. 

\begin{table}
\renewcommand{\arraystretch}{1.4}
\begin{tabular}{lccc}
\hline\hline
Decay Mode & $\ds \Rpmz_{\mathrm{PS}}$
& $\ds \Rpmz_{\mathrm{CE}}$
& $\Rpmz_{\mathrm{PS}}\, \Rpmz_{\mathrm{CE}}$\\
\hline\hline
$\Upsilon(4S)\to B\Bbar$ & 1.048  & 1.20 & 1.26 \\
\hline
$\Upsilon(5S)\to B\Bbar$ & 1.003 & 1.05 & 1.05\\
$\Upsilon(5S)\to B^*\Bbar^*$ & 1.004 & 1.06 & 1.06\\
\hline\hline
\end{tabular}
\label{tab:coulomb}
\caption{\label{tab::R0values} Relative phase space factors $\Rpmz_{\mathrm{PS}}$ for \ups{4} and \ups{5} decays, together with the naive Coulomb
enhancement for point-like particles $\Rpmz_{\mathrm{CE}}$ and their product, corresponding to the naive prediction for $\Rpmz$.}
\end{table}

While the determinations of $\Rpmz$ in Table~\ref{tab::Rsummary} can be considered robust, as they explicitly include uncertainty estimates for the assumptions made, they are still unsatisfactory in several ways:
\begin{itemize}
    \item The overall precision is not at the level necessary for high-precision measurements at current and future colliders.
    \item There is no clear path to reduce the uncertainties related to isospin breaking, so additional measurements from category~III (or $B\to X_s\gamma$, from category~II) would not reduce this uncertainty further.
    \item The average uses decay modes, whose isospin asymmetries are themselves of interest. This concerns for instance the resolved photon contributions in $B\to X_s\gamma$ or the annihilation contributions in $B\to J/\psi K$ decays. Overall, it would be desirable for applications in $B$ physics to have a determination of $\Rpmz$ that does not rely on specific $B$ meson decays, but rather only on properties of the $\Upsilon$ system. Of the methods employed so far, only the double-tag technique fulfills this criterion.
    \item The difficulty in using only the double-tag technique is that it requires very large data sets, due to the low efficiency for double-tag events, even if using a semi-inclusive tagging. 
\end{itemize}

For these reasons, having independent methods that do not rely on assumptions about specific $B$ decays would be important.
Below we propose two such methods.

\section{Determining \texorpdfstring{\Rpmz}{R+-0}  
using \texorpdfstring{$\Upsilon(5S)$}{Upsilon(5S)} decays}
\label{sec:5Sidea}

\begin{table*}[t]
\renewcommand{\arraystretch}{1.4}
\begin{tabular}{llll}\hline\hline
    & Belle  & Belle II partial & Belle II full  \\
\hline\hline
\multicolumn{1}{l}{$\mathcal{L}_{\Upsilon(5S)}$ / $\mathcal{L}_{\Upsilon(4S)}$ [ab$^{-1}$/ab$^{-1}$]} & 0.12 / 0.71  & 0.5 / 5  & 5 / 50  \\
\multicolumn{1}{l}{$N_{B^{(*)}B^{(*)}}^{\Upsilon(5S)}$  / $N_{BB}^{\Upsilon(4S)}$} & $2.74 \times 10^7$  / $7.72 \times 10^8$ & $1.13 \times 10^8$ / $5.55 \times 10^9$  &  $1.13 \times 10^9$ / $5.55 \times 10^{10}$ \\[2pt]
\hline
\multicolumn{1}{l}{$f,\ f'$} & \multicolumn{3}{c}{$\Delta r(f,f')/r(f,f')$\hspace*{3cm}} \\
\hline
\multicolumn{1}{l}{$J/\psi K^+, \ J/\psi K^0$} & 7.1\%  &  3.5\% & 1.1\% \\
\multicolumn{1}{l}{$\bar D^0\, \pi^+, \ D^-\pi^+ $} &  2.4\% & 1.2\% & 0.4\% \\
\multicolumn{1}{l}{$\bar D^{*0} \ell^+\nu, \ D^{*-}\ell^+\nu$} & 4.5\% & 2.2\%  & 0.7\% \\
\multicolumn{1}{l}{$\bar D^0 \pi^+,\ D^{*-}\ell^+ \nu $} &  1.8\% & 0.9\% & 0.3\% \\
\hline\hline
\end{tabular}
\label{tab:estimates}
\caption{Estimated sensitivity to $r(f,f')$ in Eq.~(\ref{doubleratio}), with available Belle data and anticipated partial and full Belle~II~data.}
\end{table*}

As discussed in the previous sections, the main reason for the sizeable isospin violation causing $\Rpmz$ to deviate from unity is the small phase space in $\Upsilon(4S)$ decays, $m_{\Upsilon(4S)} - 2 m_B \simeq 20 \, \mathrm{MeV}$, while
the mass difference near the $\Upsilon(5S)$ resonance is more substantial, $m_{\Upsilon(5S)} - 2 m_B \simeq 326\,\MeV$. However, an $e^+e^-$ collider running near this resonance produces many different final states.

Experimentally, $\Gamma(\Upsilon(5S) \to BB X) = (76.2^{+2.7}_{-4.0})\%$~\cite{Workman:2022ynf, Belle:2021lzm}, of which only about 5.5\,\% is direct $B\Bbar$ production, complemented by $BB^*$ (13.7\%) and $B^*B^*$ (38.1\%) production.%
\footnote{Regarding phase space differences caused by the $B^*$ masses, the mass difference $m_{B^{*0}} - m_{B^{*+}} = (0.91 \pm 0.26)\,\MeV$~\cite{CMS:2018wcx} has been measured by CMS.
Curiously, this value is approximately $-m_c/m_b \simeq -\frac13$ times $m_{D^{*0}} - m_{D^{*+}} \simeq -3.4\,\MeV$~\cite{Workman:2022ynf}, as expected from heavy-quark symmetry. The isospin splittings of the ground-state mesons, $m_{B^0} - m_{B^+} \approx 0.3\,\MeV$ and $m_{D^0} - m_{D^\pm} \approx -4.8\,\MeV$~\cite{Workman:2022ynf}, are far from this relation, probably due to electromagnetic effects.}
Additionally multi-body final states, such as $B^{(*)}B^{(*)}\pi$ and $BB\pi\pi$ contribute. For the (quasi-)two-body final states, we expect 
\beq\label{eq:5Sratio}
\Rpmz_{5S} = \frac{\Gamma(\Upsilon(5S) \to B^{(*)+} B^{(*)-})}{\Gamma(\Upsilon(5S) \to B^{(*)0} \Bbar^{(*)0})} \simeq 1 \,.
\eeq
This allows for a novel determination of $\Rpmz$, by studying 
the double ratio of pairs of decays at the $\Upsilon(4S)$ and $\Upsilon(5S)$ resonances,
\beq\label{doubleratio}
r(f,f') = \bigg[ \frac{N(B^+\to f)}{N(B^0\to f')} \bigg]_{\Upsilon(4S)}\, \bigg/
\bigg[ \frac{N(B^+\to f)}{N(B^0\to f')} \bigg]_{\Upsilon(5S)}
\, .
\eeq 
Here $N$ denotes the acceptance- and efficiency-corrected yields in $\Upsilon(4S)$ and $\Upsilon(5S)$ decays, in the latter case including $B$ mesons from all (quasi-)two-body decays $\Upsilon(5S)\to \bar B^{(*)}B^{(*)}$.
Crucially, in this ratio the ${\cal B}(B^+ \to f)$ and ${\cal B}(B^0 \to f')$ branching fractions cancel, so no information on the size of isospin breaking in the decay rates is needed. In fact, $f$ and $f'$ do not have to be (pseudo\nobreakdash-)isospin related, and any pair of states can be chosen to minimize the experimental uncertainties.
Thus, the double ratio in Eq.~(\ref{doubleratio}) directly probes the ratio of production rates, $\Rpmz$, assuming Eq.~(\ref{eq:5Sratio}) holds. 

One aspect that could spoil Eq.~(\ref{eq:5Sratio}) is the conta\-mi\-nation from final states other than $B^{(*)}\Bbar^{(*)}$,
where the reduced phase space may enhance isospin violation. 
However, if the $B \to f$ decay is reconstructed in a fully hadronic final state, its kinematic properties can be used to separate many-body from the (quasi\nobreakdash-)two-body production, using the beam-constrained mass $M_{\mathrm{bc}} = \sqrt{s/4 - |\vec p_B|^2}$~\cite{Belle:2010hoy}, 
where $\vec p_B$ is the three-momentum of the reconstructed $B$ meson. The $M_{\mathrm{bc}}$ method can also be used for semileptonic decays~\cite{BaBar:2010efp}.

In Table~\ref{tab:estimates} we present estimates of projected sensitivities to $r(f,f')$ using this method, with the existing Belle, 
as well as anticipated Belle~II data, the latter split into partial ($10\%$) and full data sets.
We studied a few promising modes, corresponding to different parton-level transitions and different experimental signatures and uncertainties. We base our uncertainty estimates on Refs.~\cite{Belle:2021udv,Belle:2022hka,BaBar:2007cke,BaBar:2007nwi} and assume for Belle~II an improvement on the systematic uncertainties by a factor of two. We scale the statistical uncertainties with the integrated luminosity ratios and base our estimates for the $\Upsilon(5S)$ analyses on the precision of the $\Upsilon(4S)$ measurements, assuming the same systematic uncertainties, but correspondingly larger statistical uncertainties. 
We further assume that common systematic uncertainties cancel between the $\Upsilon(4S)$ and $\Upsilon(5S)$ measurements. 

For $B \to J/\psi K$ decays and the currently available Belle $\Upsilon(4S)$ and $\Upsilon(5S)$ data sets, a precision of 7.1\% on $R^{\pm0}$ can be reached, limited by the statistical uncertainty of the $\Upsilon(5S)$ measurement. 
A determination focusing on $B \to D \pi^\pm$ decays could already reach a precision similar to the current world average. Semileptonic $B \to D^* \ell \bar \nu$ also offer a clean avenue, but are limited by the $B^+ \to \bar D^{*\,0} \ell^+ \nu$ precision at $\Upsilon(5S)$. An additional improvement could be obtained by focusing on $B \to D X \ell \bar\nu$ decays. 
These three decays look promising to reach $1\%$ or even sub-1\% uncertainties with a Belle~II data set of \SI{5}{\per\atto\barn} of $\Upsilon(5S)$ data.
We illustrate the fact that $f$ and $f'$ can be chosen independently to minimize the experimental uncertainties by studying mixed $D\pi$ and semileptonic channels, further improving the precision of the $r(f,f')$ determination.

We conclude that the double-ratio method using either $B \to D \pi^\pm$ decays or mixed $D\pi$ and semileptonic decays is a promising way to study the feasibility of this method with the existing Belle data.

\begin{figure*}[tb]
    \includegraphics[width=.85\textwidth, clip, bb=100 15 1175 455]{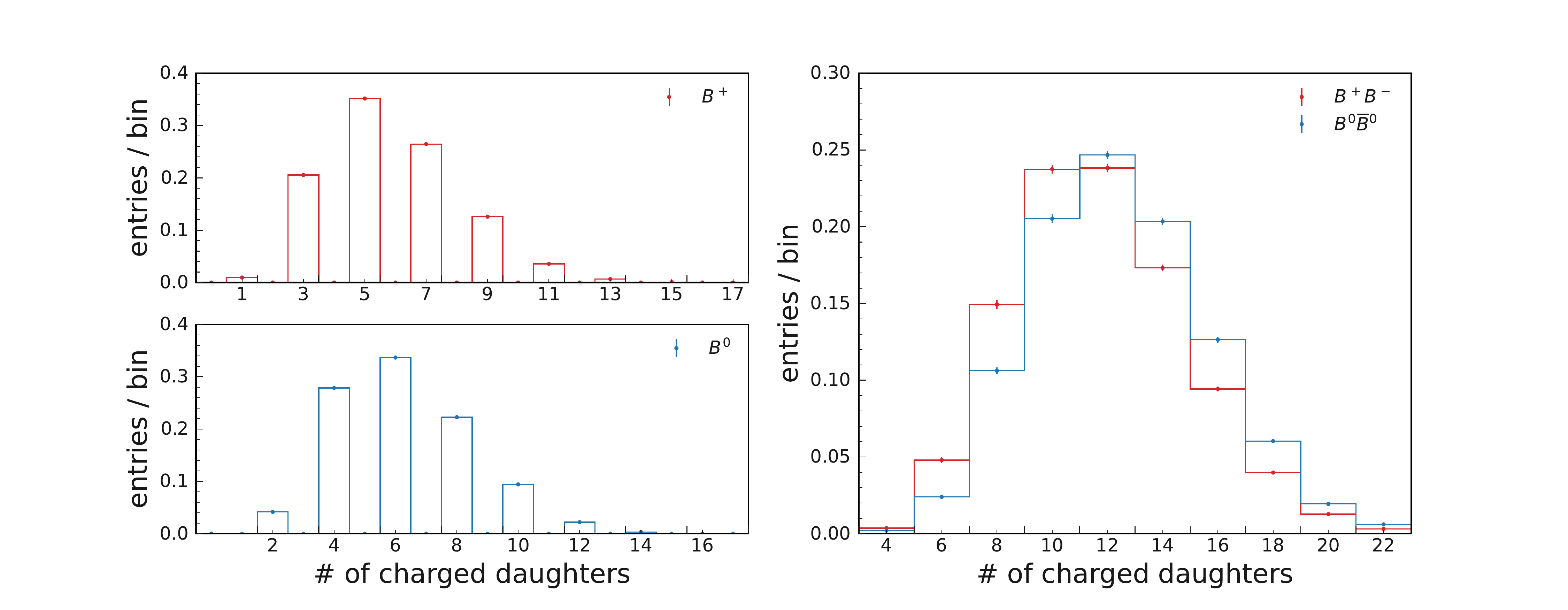}
    \caption{Number of charged daughters from $B^+$ and $B^0$ decay (left) and from a pair of $B$ decays (right), from {\tt EvtGen}.}
    \label{fig:bb_dists}
\end{figure*}

\section{Decay-channel-independent determination of \texorpdfstring{$\Rpmz$}{R+-0}}
\label{sec:tracks}

An alternative to using specific decay channels for determinations of \Rpmz\ would be the use of the full range of $B$ meson decays. This would constitute another way to remove assumptions about isospin violation. 
To our knowledge, such a determination has not been attempted; in the following 
we discuss a possible strategy and estimate the corresponding sensitivities.
This idea utilizes the fact that
$B$ mesons leave a fairly easy signature to trigger on, and that $e^+e^-$ $B$-factory experiments operate often with nearly 100\% efficiency to record events. Most triggers rely on properties that are nearly identical for $B^0\B0bar$ and $B^+B^-$ events: a typical selection requires at least three tracks and more than \SI{1}{\GeV} energy deposition in the calorimeter with four isolated clusters. Such inclusive samples are in fact regularly analyzed to count the number of $B$ meson pairs and to subtract backgrounds from continuum processes~\cite{BaBar:2014omp}.

To separate $B^0\B0bar$ and $B^+B^-$ events, another event property can be combined with this approach: the total number of detector-stable charged daughter particles. This is a difficult quantity to reconstruct, but has a reliable proxy with the total number of charged-particle tracks.  Figure~\ref{fig:bb_dists} (left; top and bottom) show the number of charged daughter particles for $B^+$ and $B^0$ meson decays as simulated by \texttt{EvtGen}~\cite{Lange:2001uf}, without any selection. 
A distinctive feature is that for $B^+$ ($B^0$) the number of charged daughters must be odd (even). This separation is reduced if one looks at the number of charged daughters of a pair of $B$ mesons produced in $\Upsilon(4S)$ decay (as shown in Fig.~\ref{fig:bb_dists}, right).
A key problem is that these distributions are sensitive to the modeling of $B$ meson decays. For instance, in \texttt{EvtGen} thousands of exclusive decays are mixed with final states from \texttt{Pythia8}~\cite{Sjostrand:2014zea} to simulate inclusive $B$ meson decays. 
One way to control this is to measure this distribution or rather its proxy (the number of charged-particle tracks) in data using
decays of $B$ mesons, which identify their charge. For instance, one can consider $B^0 \to D^- \pi^+$ decays, which, despite its small branching fraction of $\approx 2.5 \times 10^{-3}$, can be reconstructed with excellent experimental precision.

With the final-state particles of one $B$ meson decay precisely assigned to this signal, the rest of the collision event can be assessed and the multiplicity distribution of the number of charged-particle tracks can be precisely measured. Similar measurements can be carried out with $B^\pm$ decays and with other exclusive channels.  This way one can obtain the key ingredients for the prediction of the $B$ meson pair distributions from data, as their decays progress fully independently from each other. 

We construct an Asimov fit~\cite{Cowan:2010js} to assess the separation power using the number of charged particles. 
We assume that a calibration of the charged and neutral $B$ meson multiplicities can be carried out with $B^0 \to D^- \pi^+$ and $B^+ \to D^0 \pi^+$ decays. 
We scale the statistical uncertainty of Ref.~\cite{Belle:2021udv} by the expected increase in the integrated luminosity to evaluate the future Belle~II sensitivities for \SI{5}{\per\atto\barn} and \SI{50}{\per\atto\barn}. As we do not need to measure a branching fraction, but rather the distribution of reconstructed tracks, many of the leading systematic uncertainties in Ref.~\cite{Belle:2021udv} do not dilute the sensitivity. Using the number of events and the expected distributions, we determine templates and correlated uncertainties for the $B^0\B0bar$ and $B^+B^-$ multiplicity distributions. 
Notably, the predictions in the bins of the multiplicity for a pair of $B$ mesons are correlated, as they are predicted from sampling twice the distribution of a single $B$ meson decay. We fit the resulting distributions with different assumptions for the three luminosity scenarios and report the achievable relative uncertainties in Table~\ref{tab:Rpm_incl}, taking into account the uncertainties from the expected calibration precision. 

\begin{table}[tb]
\renewcommand{\arraystretch}{1.4}
\begin{tabular}{cccc}\hline\hline
    & Belle  & Belle II partial & Belle II full  \\
\hline\hline
\multicolumn{1}{c}{$\mathcal{L}_{\Upsilon(4S)}$ [ab$^{-1}$]}  &  0.71  &  5  & 50  \\
\multicolumn{1}{c}{$\Delta(\Rpmz)/\Rpmz$}  &  2.2\%  &  0.9\%  &  0.3\% \\
\hline\hline
\end{tabular} 
\label{tab:Rpm_incl}
\caption{The estimated \Rpmz sensitivity achievable using the number of charged-particle tracks. This includes the calibration uncertainty in the number of charged-particle tracks from $B^0 \to D^- \pi^+$ decays and assumes a similar sensitivity can be achieved in $B^\pm \to D^0 \pi^+$ decays. Without the calibration uncertainty, the statistical component would be sub-percent even with the data available now.}
\end{table}

In practice, additional reconstruction effects will cause  differences between the number of charged particles and the number of tracks, such as the finite detector acceptance or the occurrence of misidentified or duplicate tracks. Such effects shift and broaden the $B^0\B0bar$ and $B^+B^-$ distributions, and a more robust study on the feasibility of this method can only be done within the experiments.

\section{Production fraction ratios at hadron colliders}
\label{sec:LHC}

At hadron colliders, the production fractions of charged and neutral $B$ mesons, denoted $f_u$ and $f_d$, respectively, play an analogous role to that of the $\Upsilon$ decay rates at $e^+e^-$ $B$-factories. However, symmetry considerations are not as easily applicable: a priori, we \emph{cannot} expect the relation $f_u=f_d$ to hold, since both the initial and final states are more complicated than at a $B$-factory. At the Tevatron, the initial $p\bar p$ state is a superposition of an isosinglet and an isotriplet, while at the LHC the $pp$ initial state is a pure isotriplet. 
Furthermore, the presence of additional particles in the final state does not allow for a determination of the isospin state of the $b$ hadron pair.
However, the dominant $b\bar b$ production mechanisms at the LHC (gluon splitting and $t$-channel flavor creation) are isospin invariant.
At the same time, the fragmentation into $B$ mesons is a complicated process. 
Regarding fragmentation to $B_s$ mesons, corrections to the $SU(3)$ flavor symmetry are large, as measured by the ratio $f_s/f_d \approx 0.25$, with a dependence on the center-of-mass energy and kinematics~\cite{LHCb:2021qbv}.  
Therefore, the size of the ratio $f_u/f_d$ is ultimately an experimental question and $f_u/f_d=1$ cannot be assumed, but should be determined experimentally, including a possible kinematic dependence, as observed for $f_s$ and $f_{\Lambda_b}$~\cite{LHCb:2019fns,LHCb:2021qbv}.

The experimental determination of this quantity is again complicated by the difficulty of decoupling the production fractions from the decay rates. An additional complication arises due to the uncorrelated hadronization of the $b$ and $\bar b$ quarks produced, such that category~I measurements discussed above are not possible. This leaves us with categories~II and III. 

A measurement falling into category II with external inputs of the ratio of decay rates (and thereby \Rpmz) for $B^0 \to J/\psi K^{*0}$ and $B^+ \to J/\psi K^+$ has been carried out by the CMS Collaboration~\cite{CMS:2022wkk}, yielding
\beq
\frac{f_d}{f_u} = 1.015 \pm 0.051\,. 
\eeq
The precision of this measurement is presently limited by the uncertainty in the CMS analysis and to lesser extent by the uncertainty in $\Rpmz$.

On the other hand, it would be desirable to obtain a measurement of $f_d/f_u$ that does not rely on the external measurement of \Rpmz, using the large samples of $B$ mesons that already exist at the LHC and will be significantly enlarged in the HL-LHC era.
To that aim, we propose to use the approximate equality of rates of the semi-inclusive decays, 
\beq
\Gamma(B^0\to \bar D^{(*)}X\mu\nu)\approx \Gamma(B^+\to \bar D^{(*)}X\mu\nu)\,.
\eeq
This relation follows from the equality of inclusive rates discussed above, given the small fraction of decays that do not result in a $D^{(*)}$ meson in the final state, specifically final states including $D_s^{(*)}\bar K^{(*)}$ or baryons. An analogous method has been employed in the determination of the ratio of production fractions $f_s/f_d$ from semileptonic decays by the LHCb collaboration~\cite{LHCb:2019fns}.
These final states also include decays of $B_s$ and $\Lambda_b$.
While most of these decays have not been observed explicitly, the ones that have been seen sum to a branching fraction of $\sim 1\%$. 
Their contributions are additionally suppressed by the smaller production fractions, $f_s$ and $f_{\Lambda_b}$, respectively, so 
accounting for them should not be too difficult~\cite{LHCb:2016gsk}.
In order to separate the neutral and charged $B$ mesons decaying into these final states, one possibility is to employ the oscillations in $B^0$ meson mixing, which are absent for $B^\pm$ mesons. This has been used by the LHCb collaboration in a time-dependent semi-inclusive measurement of $\Delta m_d$~\cite{LHCb:2016gsk} to remove the background from $B^\pm$ mesons; here this background is considered instead part of the signal.

Whether the desired ${\cal O}(1\%)$ precision can be reached via this method is an experimental question; we leave the detailed studies to dedicated experimental analyses, and simply point out their potential use for measuring $B$ meson production fractions.

Finally, large samples of $t\bar t$ events accumulated at the LHC by the ATLAS and CMS Collaborations could also be used as a way to test isospin invariance in production and/or decay of $B$ mesons. Unlike the $pp$ initial state, the $t\bar t$ system is an isospin singlet. 
If the interaction with the rest of the event, often referred to as ``color reconnection'', is small, and we consider the case in which the $W$ bosons from the subsequent $t\to b\,W^+$ process decay leptonically, we can expect $f_u = f_d$ for $B$ mesons produced in this process, based on the isospin symmetry.

In this sense, the $t\bar t$ system at the LHC can play a similar role as the $\Upsilon(4S)$ or $Z$ at $e^+e^-$ colliders, as an isosinglet source of $B$ mesons. It is therefore interesting to experimentally test the $f_d = f_u$ relationship in top quark decays, and also to test the equality of $B$ and $\bar B$ production, which could be affected, e.g., by the valence quarks in the protons. 
This can be achieved by tagging the top quark (or antiquark) in the event by measuring the charge of the lepton in a leptonic $W$ boson decay from the $t \to b W^+$ or $\bar t \to \bar b W^-$ process and then compare the yield of charged and neutral $B$ mesons produced in the fragmentation of the $b$ jet accompanying the $W$ boson. 
If this ratio is different from unity, it could have a profound impact on our understanding of color reconnection~\cite{Argyropoulos:2014zoa}. 
In any case, if $f_{u,d}$ can be determined with good precision in this process, the $t\bar t$ system can be used to probe isospin invariance in $B$ meson decays. 

Again, we leave the detailed studies of the feasibility of this approach to experiments, and simply mention them as a complementary approach to test the isospin invariance and determine production fractions of $b$ hadrons using decays of top quarks, which has not been done before.

\section{Discussion and Conclusions}
\label{sec:discussion}

Before concluding, we would like to mention a few aspects regarding the experimental environment that also affect the picture discussed so far.
Throughout this paper we have assumed that \Rpmz is a constant. This would be correct if the $\Upsilon(4S)$ would itself be produced in a decay process, but at a $e^+e^-$ collider its mass is constrained by the center-of-mass-energy $\sqrt{s}$ of the colliding beams. The typical beam energy spread at $B$-factories, such as PEP-II, KEKB, or SuperKEKB, is about 4--\SI{6}{\MeV}~\cite{BaBar:2003hmt, Akai:2018mbz}, which is several times smaller than the width of the $\Upsilon(4S)$, \SI{20.5}{\MeV}~\cite{Workman:2022ynf}. 
Assuming a Gaussian distribution for the beam energy spread, the functional dependence of $\Rpmz$ on $\sqrt{s}$ results in a small bias. 
There is again a substantial range of predictions for this energy dependence~\cite{Byers:1990rd, Kaiser:2002bm, Voloshin:2003gm, Voloshin:2004nu, Dubynskiy:2007xw, Milstein:2021fnc}.
Using as examples the phase space estimate in Eq.~(\ref{eq:rp_ps}) or the simplified Coulomb factor in Eq.~(\ref{eq:rp_qed}) (which appears to be an overestimate), we find that the impact of the beam energy spread is 0.3\% or 0.4\%, respectively, which is currently an order of magnitude smaller than the experimental accuracy. We expect that this effect is much smaller at the $\Upsilon(5S)$ resonance. 

\begin{figure}[tb]
    \includegraphics[width=.85\columnwidth, clip, bb=10 0 390 255]{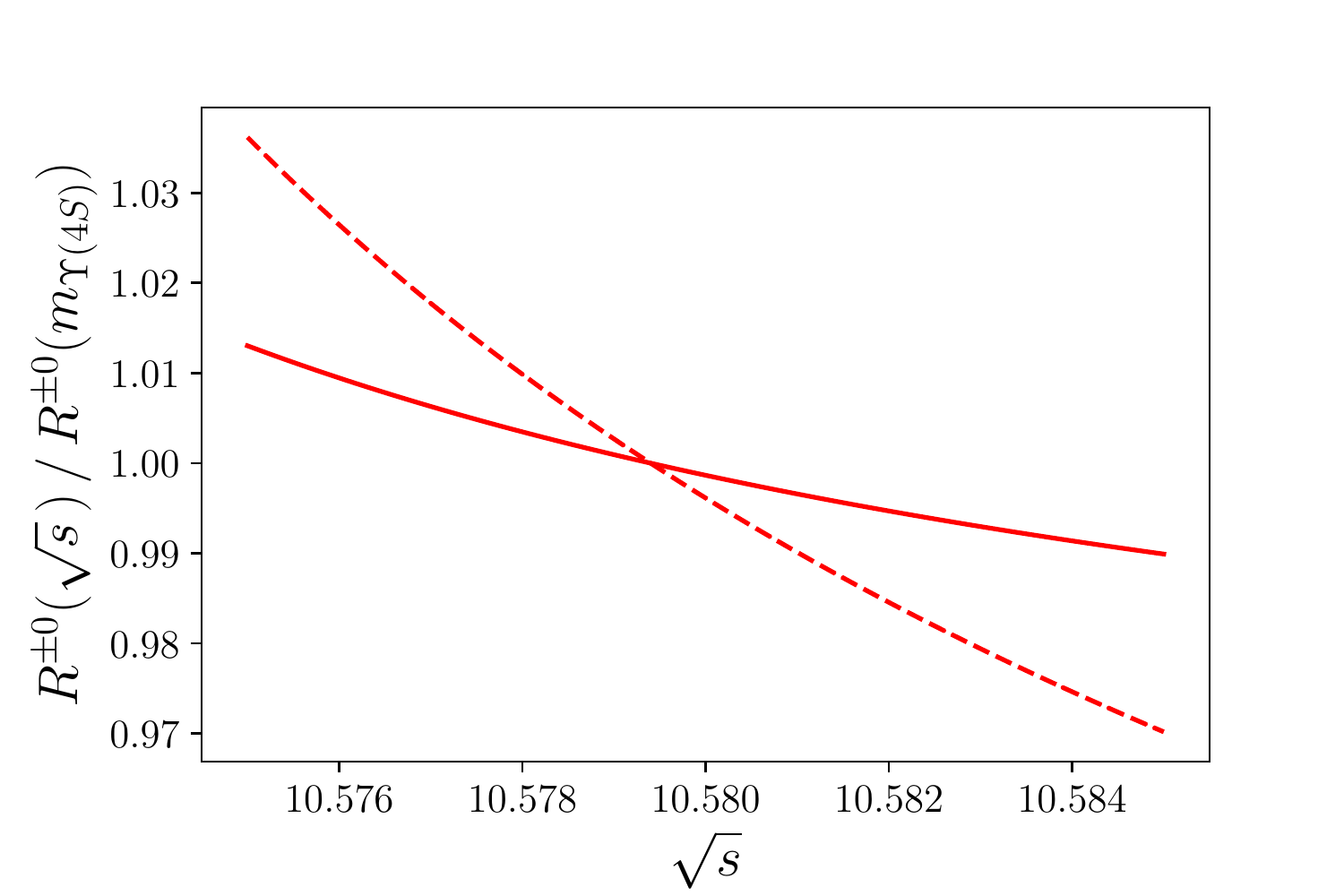}    
    \caption{The ratio \Rpmz as a function of $\sqrt{s}$ using Eq.~(\ref{eq:rp_ps}) (solid) and Eq.~(\ref{eq:rp_qed}) (dashed), normalized to the respective values at $\sqrt{s} = m_{\Upsilon(4S)} = \SI{10.5794}{\GeV}$~\cite{Workman:2022ynf}.}
    \label{fig:rp0_rp0_ratio}
\end{figure}

Another interesting question is what happens to \Rpmz if different experiments run at different center-of-mass energies, near, but not exactly on the peak cross section of $e^+ e^- \to b \bar b$ in the vicinity of the $\Upsilon(4S)$ resonance. Such shifts can also occur during different runs of a single experiment. 
If the total data of a given experiment is used to extract branching fractions of the same experiment using an identical data set, there is of course no problem: the recovered \Rpmz values correspond to the recorded data. 
However, if several experiments are combined, or within an experiment \Rpmz determinations use different data taking periods and conditions, biases may emerge. 
We can estimate possible shifts by studying Fig.~\ref{fig:rp0_rp0_ratio} qualitatively: if the phase space dependence is the leading contribution that changes \Rpmz as a function of $\sqrt{s}$, shifts of the order 1\% can occur for order \SI{5}{\MeV} shifts away from the peak cross section. 
If Eq.~(\ref{eq:rp_qed}) is used instead, these shifts can be as large as 3\%. 
Since, as discussed above, we cannot rely on any of these estimates, the energy dependence of \Rpmz
should be experimentally determined. This can be done, in a limited range, by exploiting the varying running conditions that provide samplings around the peak $e^+ e^- \to b \bar b$ cross section. For a more complete exploration, a dedicated energy scan and measurements using modes that can be reliably identified,
such as $B \to J/\psi K$, are required.

Another key question resides in the experimental determination of the number of $B$ meson pairs and its robustness against \fnB $\neq 0$. 
In order for Belle~II to achieve percent-level precision goals in the study of branching fractions and other observables, a consistent treatment is needed that takes into account the correlated aspects of \Rpmz determinations and $B$ meson counting. 

We investigated the determinations of the $\Upsilon(4S)\to B^+B^-$ and $B^0\B0bar$ decay rates and proposed new methods to improve them.  
Presently the limited precision of these decay rates constitutes a lower limit of $\sim2\%$ on the uncertainties in absolute branching fraction measurements, and thereby in applications, such as precision determinations of CKM matrix elements or flavor symmetry relations.
We revisited the theoretical assumptions in \Rpmz measurements, and updated its world average in Sec.~\ref{sec:reapp}, emphasizing underestimated uncertainties in prior evaluations, in particular due to isospin violation and non-zero \fnB value (as shown in Fig.~\ref{fig:f00fpm}).  Due to the inclusion of additional measurements, we obtained nevertheless an improved precision for \Rpmz and the individual production fractions of about $2\%$ and $1.5\%$, given in Eqs.~\eqref{eq:Rpmzfull} and \eqref{eq:fpm00full}, respectively. When using both $f_\pm$ and $f_{00}$, care must be taken to include their correlations shown in Fig.~\ref{fig:f00fpm}.
We proposed two new methods in Secs.~\ref{sec:5Sidea} and \ref{sec:tracks} to determine \Rpmz precisely, using $\Upsilon(5S)$ data anticipated at Belle~II over the next decade, or using
the different average number of charged-particle tracks between charged and neutral $B$ meson decays.  
Tables~\ref{tab:estimates} and \ref{tab:Rpm_incl} summarize our estimates of future sensitivities.
Section~\ref{sec:LHC} proposed possible measurements of $f_d / f_u$ at the (HL-)LHC.

The issues raised and methods developed in this article will remain important at future colliders. In the meantime, progress could already be made by revisiting the measurement of Ref.~\cite{Belle:2002lms} with the full data set, performing a double-tag analysis at Belle or Belle~II, and by using our methods with the existing data.

\newpage
\acknowledgments

We thank Paolo Gambino, Dean Robinson, Frank Tackmann, and Kerstin Tackmann for helpful conversations.
We thank the organizers of the workshop on ``Challenges in Semileptonic $B$ Decays'' for the stimulating Barolo ambiance when this work started. 
We also thank the CERN theory group for hospitality. 
FB and MK are supported by DFG Emmy-Noether Grant No.\ BE 6075/1-1 and BMBF Grant No.\ 05H21PDKBA. FB also thanks the LBNL theory group for its hospitality. 
The work of MJ is supported by the Italian Ministry of Research (MIUR) under grant PRIN 20172LNEEZ.
The work of GL and ZL was supported in part by the Office of High Energy Physics of the U.S.\ Department of Energy under contracts DE-SC0010010 and DE-AC02-05CH11231, respectively.

\bibliography{refs}

\end{document}